# The Body–Algorithm–Emotion Loop: Making Emotion AI Misrecognition Visible through Artistic Practice

**Chunyi Shen**
London College of Communication
University of the Arts London
London, United Kingdom
c.shen0620251@arts.ac.uk

## ABSTRACT

This pictorial presents The Body in Algorithmic Colonialism, a screen-based interactive installation that critically examines Emotion-AI through artistic practice. The work uses facial tracking to simulate how commercial Emotion-AI systems translate facial movement into emotional categories and verdicts of compliance or non-compliance. Rather than treating explainability as technical transparency alone, the installation approaches XAI as an embodied encounter with misrecognition: viewers experience how the body is measured, classified, and returned to itself through algorithmic thresholds. Drawing on critical AI studies, affect theory, and practice-as-research, the pictorial positions artistic practice as a method for making affective classification visible and contestable. It documents the conceptual framework, technical process, and critical context of a prototype developed in the first year of a practice-based PhD. No audience responses, interviews, or participant data are analysed in this submission.

### Authors Keywords

Emotion AI; Algorithmic colonialism; Interactive installation; Affective computing; Critical AI studies; Embodied interaction; Practice-based research

## INTRODUCTION

Emotion-AI systems — technologies that claim to detect, classify, and respond to human emotions — are increasingly embedded in high-stakes decisions: job interviews, classroom monitoring, airport security, and automotive safety. Facial emotion recognition (FER), one prominent branch, maps facial movements onto emotion categories, drawing on contested psychological theories of universal expression [1, 2].

This work intervenes at the intersection of critical AI studies and artistic practice. Rather than critiquing Emotion-AI abstractly, it builds a system that enacts the logic of emotional classification in real time — making the experience of being measured, categorised, and judged perceptible through direct bodily encounter.

The installation uses TouchDesigner and MediaPipe blendshape tracking: it measures the curvature of the viewer's smile, compares it against a threshold, and delivers a verdict. The system does not ask what you feel. It asks whether you comply.

## RESEARCH CONTEXT

This installation is the first of three artistic experiments in a practice-based PhD at London College of Communication, UAL. The research examines how Emotion-AI enacts algorithmic colonialism extending colonial logics of universality and legibility into affective life through physiological and psychological discipline [3, 4].

Theoretically, the work draws on:

• Couldry & Mejias [3]: data colonialism as the extraction of human life through data

• Foucault [5]: disciplinary power operating through normalisation of the body

• Barrett [1]: emotions are constructed, not discovered — not universal

• Glickman & Sharot [6]: repeated human-AI interaction reshapes human emotional perception

The installation was developed and exhibited as part of Unfolding Narratives at London College of Communication, UAL, in 2026.

# THE INSTALLATION: SMILE AS ALGORITHMIC TEST

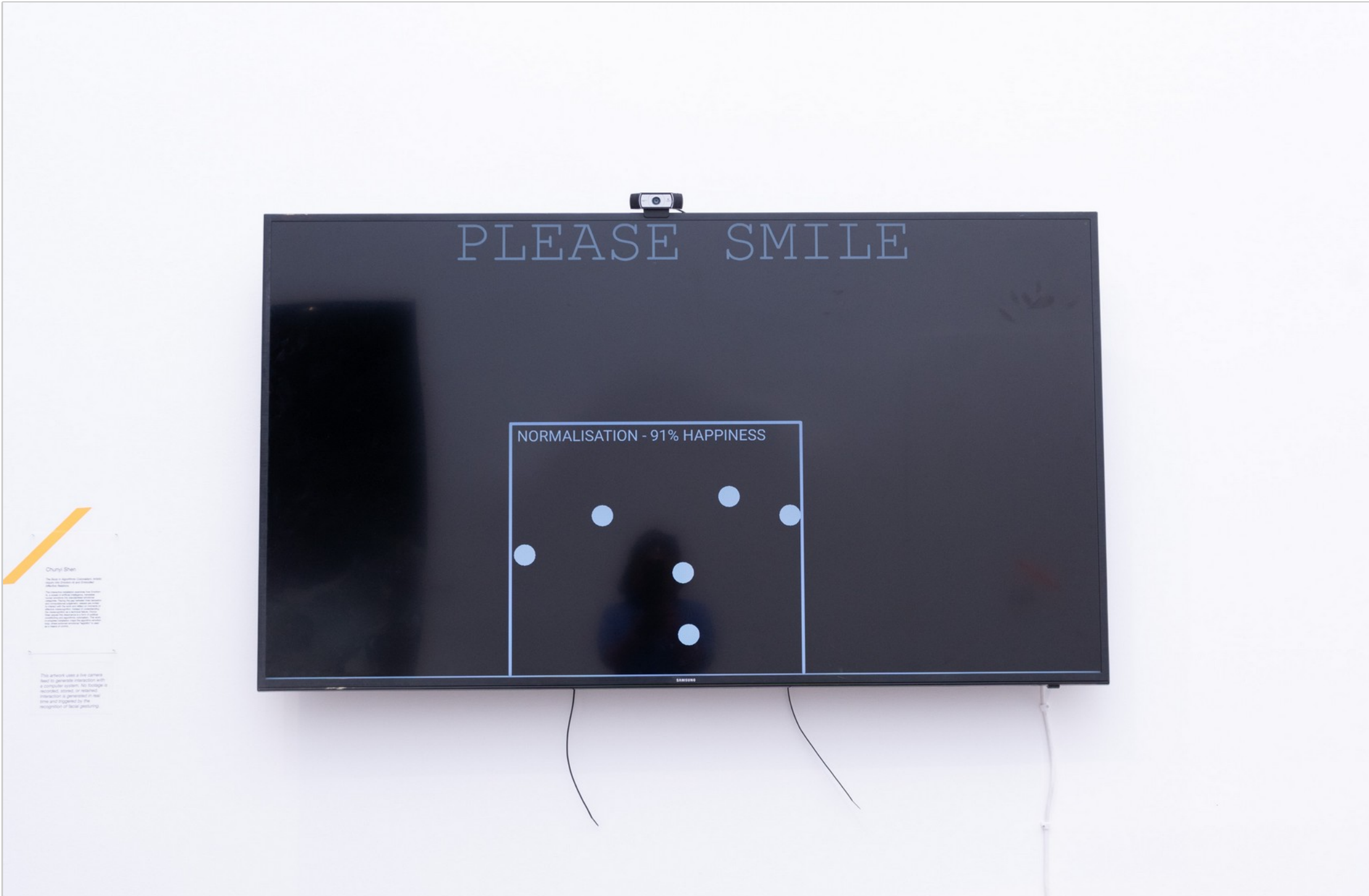

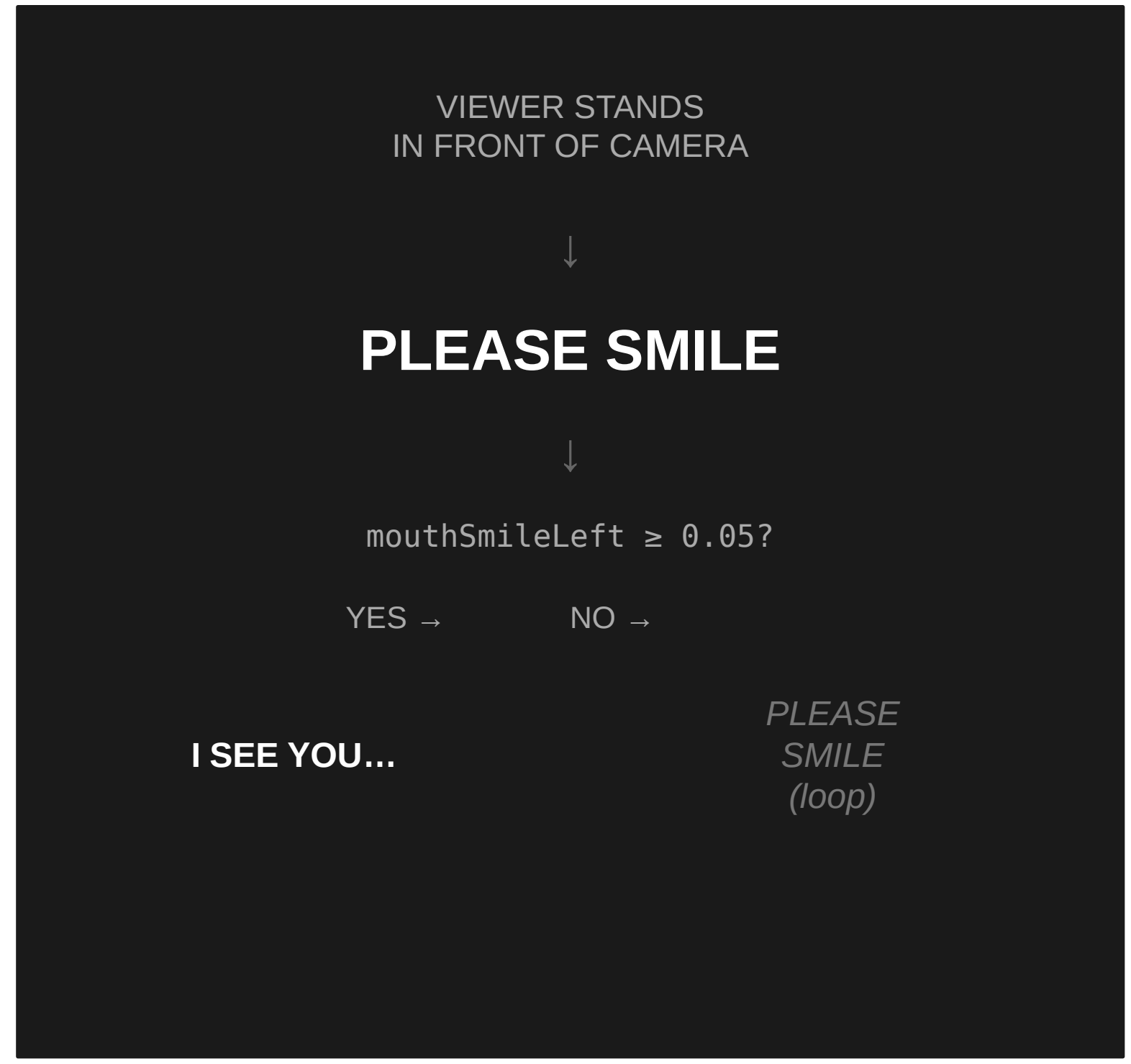

*Figure 1. Core interaction logic of the installation. The system measures facial muscle movement, not emotion — yet delivers an emotional verdict.*

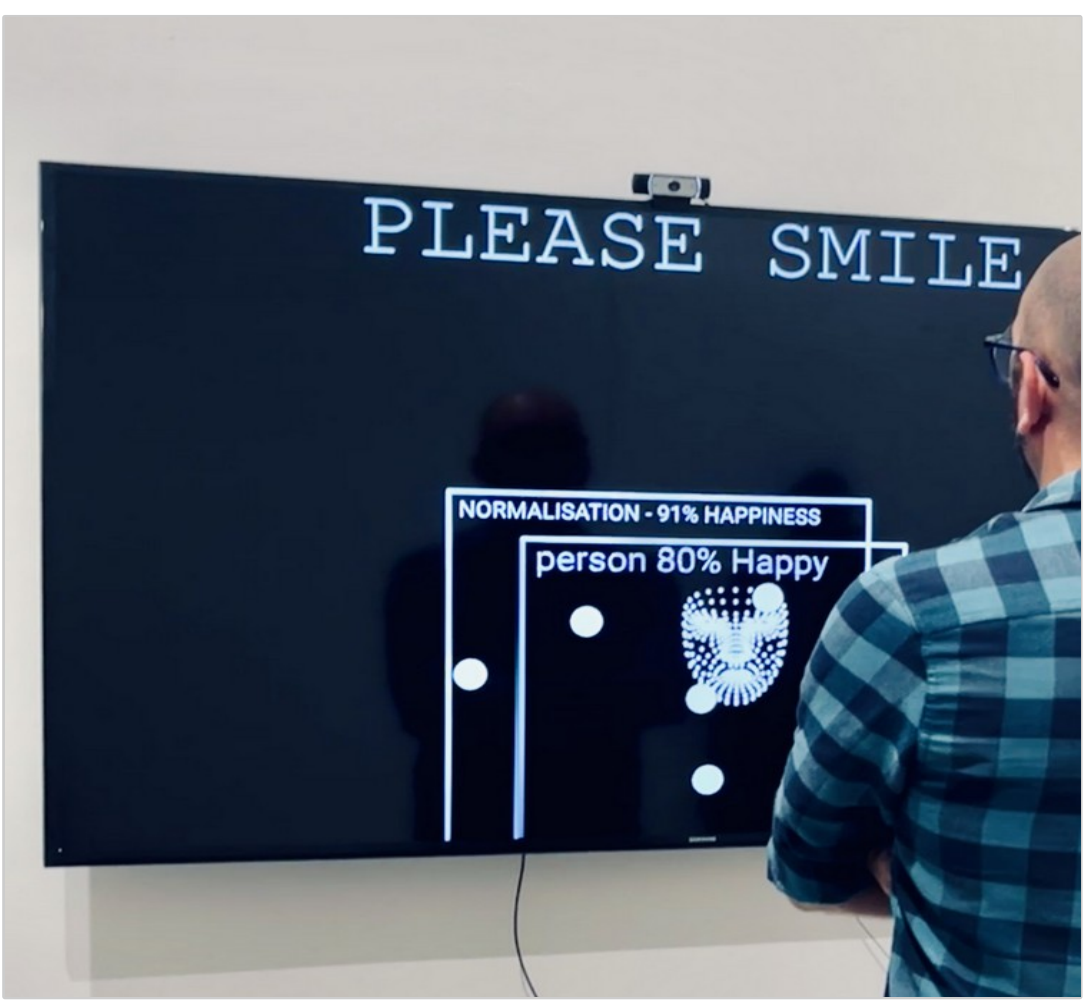

*Figure 2. 'PLEASE SMILE' — the system's demand.*

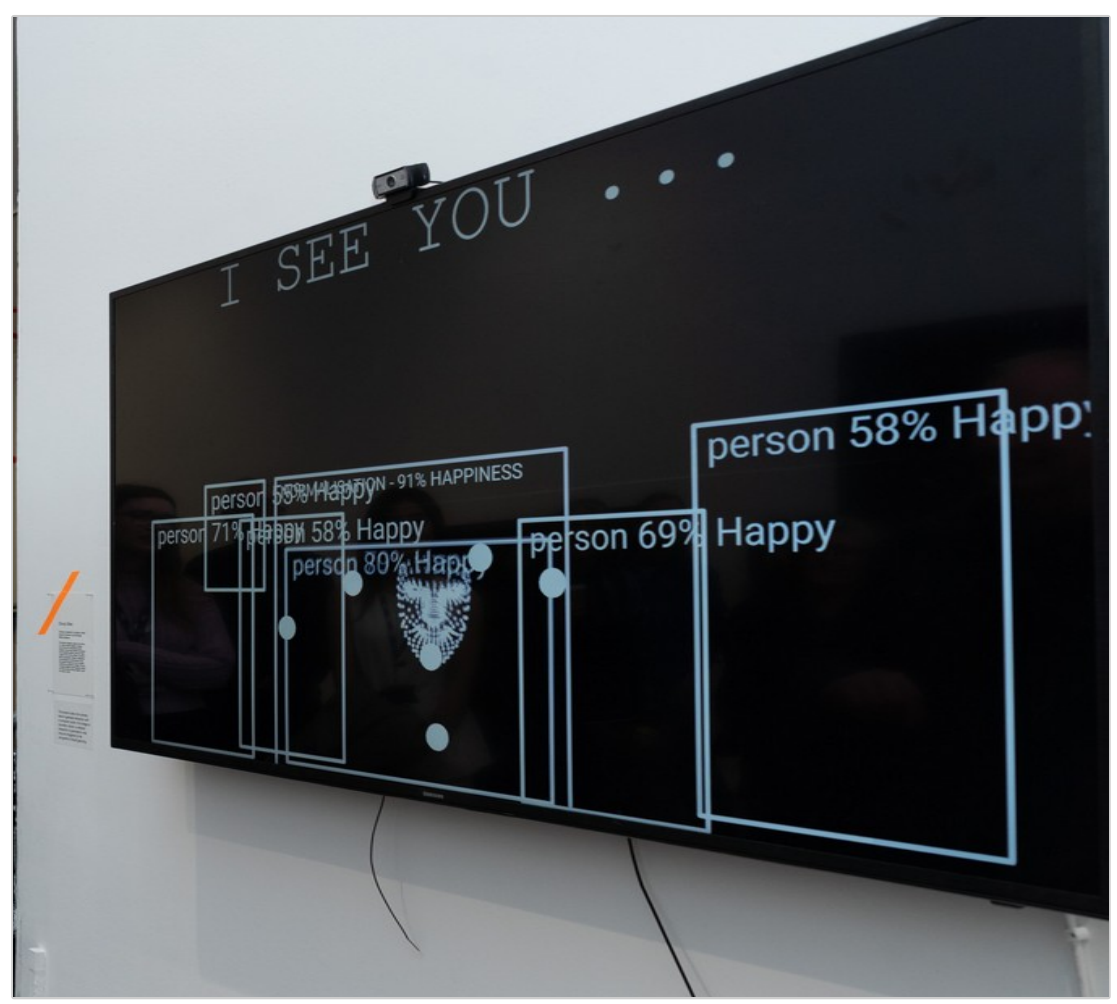

*Figure 3. 'I SEE YOU…' — compliance acknowledged.*

The installation was built in TouchDesigner with MediaPipe, which tracks facial landmarks and estimates blendshape coefficients. Rather than training an emotion classifier, the work stages the classificatory logic at its most legible layer: a manually set threshold between 'smiling enough' and 'not smiling enough', returned as a verdict rather than a validated inference of emotion.

The labels on screen are part of the work. The boxes in Figures 2 and 3 come from MediaPipe's object detector: the percentage is its confidence that a person is present, to which the installation appends the word 'Happy'. 'NORMALISATION · 91% HAPPINESS' is a fixed caption. A measure of presence is displayed as a measure of emotion.

# TECHNICAL PROCESS & THEORETICAL FRAMEWORK

## FER in Three Conceptual Layers

LAYER 01

### Ekman Classification

6 universal emotion categories
(happiness, sadness, anger,
fear, disgust, surprise)

↓

LAYER 02

### Facial Action Coding (FACS)

Muscle movements coded
as Action Units (AUs).
AU6 = cheek raiser, AU12 = lip corner puller

↓

LAYER 03

### Machine Learning Model

Trained to recognise AUs
and output emotion label
+ confidence score

*Treating emotion categories as universal carries that assumption into every reading of the face.*

```
// MediaPipe blendshape output

mouthSmileLeft      0.82
mouthSmileRight     0.79
browDownLeft        0.01
eyeSquintLeft       0.44

→ Classification:   HAPPY
→ Confidence:       91%
```

*Figure 4. Illustrative blendshape output (representative values). The classification and confidence lines are staged, not model outputs: a physical measurement, renamed as an emotion.*

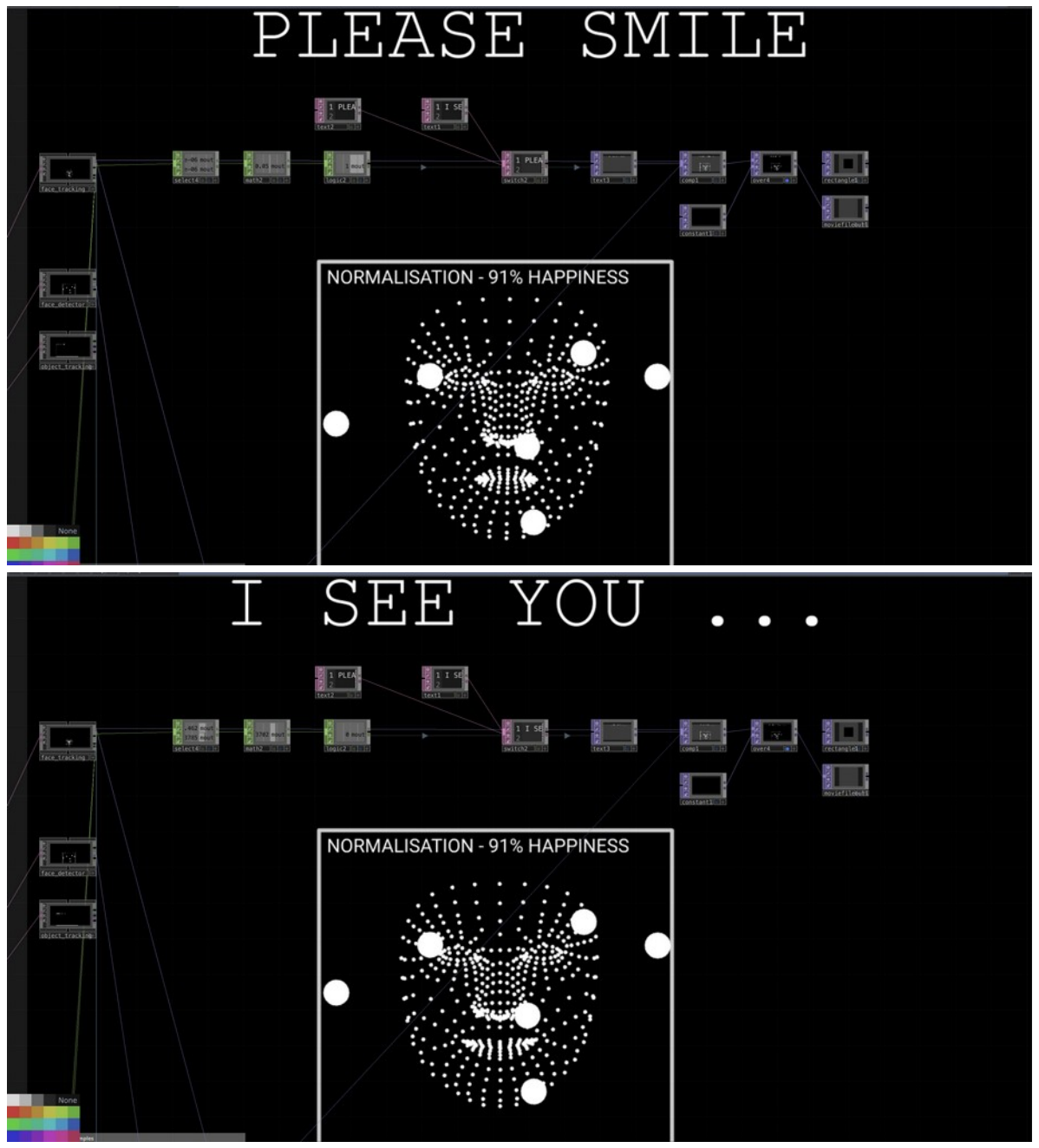

*Figure 5. The TouchDesigner patch behind the installation, tracking facial landmarks in real time. The system reads muscle curvature, not emotion.*

## The Problem of Explainability

Commercial Emotion-AI systems are largely opaque. Users cannot see the blendshape values being measured, the classification thresholds being applied, or the cultural assumptions encoded in training data. The installation makes this process perceptible — exposing the threshold behind the verdict — as a form of critical XAI: explainability not in service of efficiency, but as a tool of critique.

This aligns with what Alfrink et al. [7] call 'contestability': the capacity of those subject to algorithmic decisions to understand and challenge them.

## Theoretical Framework

Couldry & Mejias (2019) [3]

### Data Colonialism

Extraction and appropriation of human life through data — extended here into affective life.

Foucault (1977) [5]

### Disciplinary Power

Normalisation operating through the body. The installation enacts the panopticon: you smile because the system requires it.

McStay (2018) [8] · Puar (2017) [4]

### Affective Norms

Emotion-AI encodes particular standards of legibility — deciding which bodies' emotions count.

Glickman & Sharot (2025) [6]

### Body-Algorithm-Emotion Loop

Repeated AI feedback reshapes human emotional perception — the algorithm does not just read, it reshapes.

# RESEARCH QUESTIONS, CONTRIBUTION & DISCUSSION

RQ 01

How does Emotion-AI enact algorithmic colonialism through physiological and psychological discipline?

RQ 02

How can artistic inquiry be used as a research method to investigate this embodied process?

### Three Questions the Installation Raises:

***"Whose smile is the standard smile?"***

***"Whose emotion is the legible emotion?"***

***"Who decided what happiness looks like?"***

***The installation does not answer these questions. It makes you feel them.***

### Contribution & Discussion

This work makes three contributions to the XAIxArts community. First, it demonstrates artistic practice as a form of critical XAI: making algorithmic logic legible not through technical transparency alone, but through embodied encounter. Second, it applies decolonial frameworks to explainable AI, asking whose standards are encoded and whose bodies are made illegible by these systems. Third, it proposes that artistic installations can function as research instruments by staging experiential conditions that later support ethically approved qualitative inquiry.

The installation is deliberately incomplete. The current prototype simulates Emotion-AI logic without training a custom model. This is a deliberate methodological choice, though it limits how far the work speaks to commercial systems. The goal is to expose the classificatory logic at its most visible layer, before the opacity of training data and model architecture obscures it. Future iterations will compare how different systems misread the same face. The prototype stages one possible turn of a body–algorithm–emotion loop; whether repeated encounters change emotional self-perception remains an open empirical question.

*Note on current stage: This prototype, developed in the first year of a practice-based PhD, is a proof of concept. The installation demonstrates the conceptual and interactional logic of the work. Participant testing, iterative development, and integration of commercial Emotion-AI systems are planned for subsequent phases. No audience responses, interviews, or participant data are analysed in this submission. Future participant testing will be conducted only after institutional ethics approval.*